\documentclass[aps,amsmath,amssymb,superscriptaddress,
prl,showkeys,array]{revtex4}
\usepackage[dvips]{graphicx}

\newcommand{\be}{\begin{equation}}
\newcommand{\e}{\end{equation}}
\newcommand{\beml}{\begin{subequations}}
\newcommand{\eml}{\end{subequations}}
\newcommand{\beq}{\begin{eqnarray}}
\newcommand{\eq}{\end{eqnarray}}
\newcommand{\ba}{\begin{array}}
\newcommand{\ea}{\end{array}}

\begin{document}
\title{Charge and spin conductance through a side-coupled quantum dot}
\author{M. E. Torio}
\address{Instituto de
F\'{\i}sica Rosario, CONICET-UNR, Bv 27 de febrero 210bis, 2000 Rosario, Argentina}
\author{K. Hallberg}
\address{Centro At\'{o}mico Bariloche and Instituto Balseiro, 8400 Bariloche,
Argentina}
\author{C. Proetto}
\address{Centro At\'{o}mico Bariloche and Instituto Balseiro, 8400 Bariloche,
Argentina}


\begin{abstract}
The zero-temperature magnetic field-dependent conductance of electrons
through a one-dimensional non-interacting tight-binding chain with an
interacting {\it side} dot is reviewed and analized further. When the number of
electrons in the
dot is odd, and the Kondo effect sets in at the impurity site, the
conductance develops a wide minimum as a function of the gate voltage,
being zero at the unitary limit. Application of a magnetic field
progressively destroys the Kondo effect and, accordingly, the conductance
develops pairs of dips separated by $U$, where $U$ is the repulsion between
two electrons at the impurity site. Each
one of the two dips in the conductance corresponds to a perfect spin polarized
transmission, opening the possibility for an optimum spin filter. The
results are discussed in terms of Fano resonances between two interfering
transmission channels, applied to recent experimental results, and compared
with results corresponding to the standard substitutional configuration,
where the dot is at the central site of the non-interacting chain.
\end{abstract}

\maketitle


\section{Introduction}

In recent decades the electric transport through quantum dots (QD) has been
extensively studied both theoretically and experimentally \cite
{mesoscopic1,mesoscopic2,mesoscopic3}. As the result a comprehensive picture
of a big variety of underlying physical phenomena has emerged (See e.g. \cite
{Areview,ABGreview} and references therein). As QD's are small droplets of
electrons confined in the three spatial directions, energy and charge
quantization, and strong Coulomb repulsion among the electrons results from
this confinement. As these features are shared with real atomic systems,
from the very beginning an extremely useful analogy has been exploited
between ``real'' and ``artificial'' atomic systems. This analogy received
strong support through an experimental breakthrough where the Kondo effect
in quantum dots was unambiguously measured.\cite{david,sara,schmid}
Historically, the Kondo effect was introduced more than forty years ago to
explain the resistivity minimum for decreasing temperatures observed in
metallic matrices with a minute fraction of magnetic impurities.\cite{jun}
According to the detailed microscopic theory, when the temperature $T$
decreases below the Kondo temperature $T_{K}$, each of the localized
magnetic impurities starts to interact strongly with the surrounding
electronic cloud, which finally results in a singlet many-body ground state,
reaching its maximum strength at $T=0.$\cite{hewson} The minimum in the
resistivity results from the fact that, as the temperature is lowered, the
scattering with phonons decreases down to a temperature at which the
scattering with localized impurities becomes important as the Kondo effect
starts being operative. It is important to emphasize that in this case, the
so-called traditional Kondo effect, magnetic impurities act as scattering
centers, {\it increasing} the sample resistivity.

The opposite behavior is found in some circumstances in the Kondo effect in
quantum dots. The situation considered almost without exception both
theoretically and experimentally, consists of a quantum dot connected to two
leads in such a way that electrons transmitted from one electrode to other
should necessarily pass through the quantum dot ({\it a substitutional dot}%
). As theoretical calculations predicted,\cite{glazman,patrick} in this
configuration and when the dot is occupied by an odd number of electrons,
the conductance {\it increases} when decreasing the temperature and the
Kondo effect sets in, essentially due to a resonant transmission through the
so-called Kondo resonance which appears in the local density of states at
the dot site at the Fermi level of the leads. In this situation, at $T=0$,
the conductance should take the limiting value $2e^{2}/h,$ corresponding to
the unitary limit of a one-dimensional perfect transmission channel.\cite
{wiel} At finite temperatures $T>T_{K}$ the Kondo effect is destroyed and only
the standard Coulomb blockade remains\cite{KMMTWW,AL}, which manifests
itself in strong oscillations of the dot's conductance versus the gate
voltage applied to the dot.

The aim of this work is to theoretically review the results for an
alternative configuration of a {\it side-coupled} quantum dot, attached to a
perfect quantum wire. In this case, the quantum dot acts as a scattering
center for transmission through the chain, in close analogy with the
traditional Kondo effect. We have found that when the dot provides a
resonant energy for scattering, the conductance has a sharp decrease,
reminiscent of the Fano resonances observed in scanning tunneling microscope
experiments for magnetic atoms on a metallic substrate.\cite{li} and QD's coupled
sideways to a quantum wire\cite{koba}.
A similar problem has been discussed by other authors\cite{kang,aligia,franco}. We
differ from
them in that they have used approximate methods to describe the Kondo
regime (slave-boson mean field theory\cite{kang}, interpolative perturbation
scheme\cite{aligia} and X-slave-boson treatment\cite{franco}),while our numerical
results take
into account the many body interactions in a very precise and controlled
way. In addition, in Ref. \cite{kang} they use the unrealistic limit of infinite
Coulomb
repulsion for electrons inside the dot $($Hubbard $U\rightarrow \infty )$,
loosing key features as the double dip structures of Fig. 8a, which give rise
to the diamond-shaped features in differential conductance experiments.\cite
{gores} Also in recent years, several authors have studied the possibility
of spin polarization in transport through substitutional QD's \cite
{costi,recher,ohno} and other nanoscopic configurations like rings
\cite{popp}, T-shaped spin filters \cite{kiselev} and dots connected to
Luttinger liquid leads\cite{schmeltzer03}. We discuss here that if the spin
degeneracy is lifted, say, by an applied magnetic field, and if the splitting
exceeds the temperature, the resonances for spin up and spin down electrons
become well separated in the absence of the Kondo effect for the case of
side-coupled quantum dots. This gives rise to the resonant spin polarized
conductance due to the effect of Fano interference. This review is mainly
based in results presented by the authors in Refs.\cite{torio1} and \cite
{torio2}. \newline

\section{The model}

The models employed in the calculation are schematically shown in Fig. 1.
Case a), corresponding to the {\it substitutional} dot situation, consists
of two semi-infinite non-interacting tight-binding chains connected to a
central site (the dot). Case b), corresponding to the {\it side} dot,
consists of a quantum wire coupled sideways to a QD. In both cases the dot
is modeled as an Anderson impurity\cite{anderson}. For the linear response
situation, $\mu _{L}\rightarrow 0^{+}$ and $\mu _{R}\rightarrow 0^{-}$, the
Hamiltonian reads:
\begin{equation}
H=H_{0}+H_{int},
\end{equation}
where $H_{0}$ is the Hamiltonian of two semi-infinite chains with site
energies all set equal to zero,
\begin{equation}
H_{0}=-\sum_{j,\sigma }t_{j}(c_{j\sigma }^{\dagger }c_{j+1\sigma }+\text{h.c.%
}),  \eqnum{2a}
\end{equation}
with $j$ running over all chain sites. Here $c_{j\sigma }^{\dagger }$ $%
(c_{j\sigma })$ represents a creation (destruction) operator for an electron
at site $j$ with spin $\sigma ,$ and $t_{j}$ are the hopping amplitudes
between first-neighbour sites. $H_{int}$ can be written as
\begin{equation}
H_{int}^{a}=\sum_{\sigma }\left( \varepsilon _{\sigma }+\frac{U}{2}\text{ }%
n_{0\overline{\sigma }}\right) n_{0\sigma },  \eqnum{2b}
\end{equation}
for the {\it substitutional} dot configuration, while
\begin{equation}
H_{int}^{b}=\sum_{\sigma }\left[ -t_{i}(c_{0\sigma }^{\dagger }c_{\sigma }+%
\text{h.c.})+\varepsilon _{\sigma }\text{ }c_{\sigma }^{\dagger }c_{\sigma }+%
\frac{U}{2}\text{ }n_{\sigma }n_{\overline{\sigma }}\right] ,  \eqnum{2c}
\end{equation}
for the {\it side} dot configuration. In the equations above, $\varepsilon
_{\sigma }$ represents the energy at the dot site, $U$ $(>0)$ is the Coulomb
repulsion energy which penalizes the double occupancy of the dot site, and $%
t_{j}=t$ for $j\eqslantless -2$, $j\geqslant 1,$ while $t_{-1}=t_{0}=t^{%
\prime }$, $n_{0\sigma }=c_{0\sigma }^{\dagger }c_{0\sigma }$ and $n_{\sigma
}=c_{\sigma }^{\dagger }c_{\sigma }$. As we are also interested in the
behavior in a magnetic field, we consider the local energies as $\varepsilon
_{\uparrow }=\varepsilon _{0}+\Delta /2,$ $\varepsilon _{\downarrow
}=\varepsilon _{0}-\Delta /2,$ with $\Delta =g\mu _{B}H$ the Zeeman
splitting of the localized orbital, {\it i.e.} the principal magnetic
field effect is to shift the local QD levels.\cite{meir2} In the usual
experimental setup, $\varepsilon _{0}$ can be controlled through application
of a gate voltage. It is interesting to point out that both models could be
mapped to a single semi-infinite chain, with the dot sitting at the free end,
and the remaining sites corresponding to the even basis states that couples
to the dot. Both models become exactly equivalent from the point
of view of their equilibrium properties,\cite{schlott} if the hoppings are
related as follows: $t_{i}=\sqrt{2}t^{\prime }=t.$ However, the {\it transport}
properties of both models are completely different. We will set $t=t_i1$.


\begin{figure}[tbd]
\includegraphics[width=0.5\textwidth]{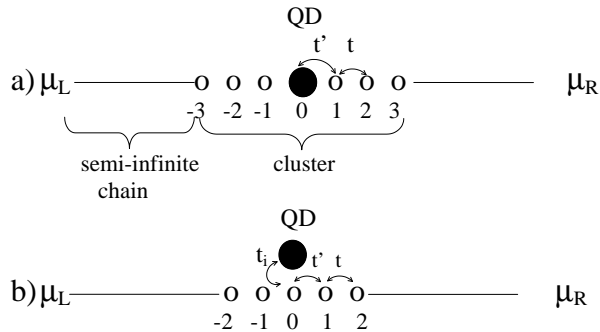}
\caption{Schematic representation of our models: a) {\it substitutional} dot
configuration; b) {\it side} dot configuration. Small open circles represent
cluster non-interacting sites, big full circle represent the dot site. The
cluster includes the dot and a few non-interacting sites. Left and right
full lines represent the non-interacting tight-binding semi-infinite chains with
their respective chemical potentials $\mu_L$ and $\mu_R$.}
\end{figure}

For the analysis of our transport results we have used the following
equation for the magnetic field dependent conductance,\cite{meir}
\begin{equation}
g(H)=\sum_{\sigma }g_{\sigma }(H)=\frac{e^{2}}{h}\frac{2\pi t^{\prime }{}^{2}%
}{t}\sum_{\sigma }\rho _{\sigma }(\varepsilon _{F})  \eqnum{3}
\end{equation}
where $\rho _{\sigma }(\varepsilon _{F})$ is the local density of states
(per spin) at site $0$ evaluated at the Fermi energy $\varepsilon _{F}.$ To
obtain the density of states $\rho _{\sigma }(\varepsilon _{F})$ in a
general interacting case $(U\neq 0)$ we use a combined method. In the first
place we consider an open finite cluster of $N$ sites in the projected space
($N=7$ in our case)
which includes the impurity. This is diagonalized using the exact
diagonalization Lanczos technique\cite{lanczos}. We then proceed to embed
the cluster in an external reservoir of electrons, which fixes the Fermi
level of the system, attaching two semi-infinite leads to its right and left%
\cite{ferrari}. This is done by calculating the one-particle Green function $%
\hat{G}$ of the whole system within the chain approximation of a cumulant
expansion\cite{cumulant} for the dressed propagators. This leads to the
Dyson equation $\hat{G}=\hat{G}\hat{g}+\hat{T}\hat{G}$, where $\hat{g}$ is
the cluster Green function obtained by the Lanczos method. Following Ref.
\cite{ferrari}, the charge fluctuation inside the cluster is taken into
account by writing $\hat{g}$ as a combination of $n$ and $n+1$ particles
with weights $p$ and $1-p$ respectively: $\hat{g}=p\hat{g}_{n}+(1-p)\hat{g}%
_{n+1}$. The total charge of the cluster $Q_{c}$ and $p$ are calculated by
solving selfconsistently the equations:
\begin{eqnarray}
Q_{c} &=&np+(n+1)(1-p),  \eqnum{4} \\
Q_{c} &=&-\frac{1}{\pi }\int_{-\infty }^{\varepsilon _{F}}\sum_{i}\text{Im }%
g_{ii}(\omega )\text{ }d\omega ,  \eqnum{5}
\end{eqnarray}
where $i$ runs on the cluster sites. Once convergence is reached, the
density of states is obtained from $\hat{G}$. It is important to stress that
this method is reliable only if $t^{\prime }$ (substitutional dot) or $t_{i}$
(side dot) are large enough, so that the Kondo cloud is about the size of
the cluster (for the parameters used here we estimate a Kondo correlation
lenght of about 10 lattice sites). Moreover, the fact that the conductance
reaches the unitary limit for the symmetric case provides an important test
of the validity of the method (see Fig. 9).

\section{Single particle $(U=0)$ Fano resonance}

Let us consider first the case $U=0,$ where an analytical and instructive
calculation of $g_{\sigma }$ is feasible, along the following lines. For
this non-interacting situation, the conductance per spin channel $g_{\sigma
} $ is linked to the transmission amplitude $\tau _{\sigma }$ by the
Landauer formula
\begin{equation}
g_{\sigma }=\frac{e^{2}}{h}|\tau _{\sigma }|^{2}.  \eqnum{6}
\end{equation}
Comparison with Eq.(5) above yields a Fisher-Lee type of relation $\left[
2\pi (t^{\prime })^{2}/t\right] \rho _{\sigma }(\varepsilon _{F})=|\tau
_{\sigma }|^{2}.$\cite{FL} For the side-dot configuration of Fig. 1b the
transmission can be computed within the one-particle picture for an electron
moving at the energy $\varepsilon $ :
\begin{eqnarray}
-\varepsilon \phi _{j,\sigma }=t(\phi _{j-1,\sigma }+\phi
_{j+1,\sigma })+ (t^{\prime }-t)(\delta _{j,1}+ \delta
_{j,-1})\phi _{0,\sigma },\hspace{0.5cm }j\neq 0 \label{gs}
\end{eqnarray}
\begin{equation}
-\varepsilon \phi _{0,\sigma }=t^{\prime }(\phi _{-1,\sigma }+\phi
_{+1,\sigma })+t_{i}\varphi _{\sigma },  \eqnum{7b}
\end{equation}
\begin{equation}
-\varepsilon \varphi _{\sigma }=-\varepsilon _{\sigma }\varphi _{\sigma
}+t_{i}\phi _{0,\sigma },  \eqnum{7c}
\end{equation}
where $\delta _{i,j}$ is the Kronecker symbol, $\phi _{j,\sigma }$ refers to
the spin-dependent amplitude of a single particle at site $j$ in the
conducting channel, and $\varphi _{\sigma }$ is the amplitude at the side
dot. The scattering problem is solved by the ansatz
\begin{equation}
\phi _{j<\text{ }0,\sigma }=e^{iqj}+\rho _{\sigma }\text{ }e^{-iqj},\qquad
\phi _{j>\text{ }0,\sigma }=\tau _{\sigma }\text{ }e^{iqj},  \eqnum{8}
\label{eq3}
\end{equation}
with the usual dispersion relation $\varepsilon =-2t\cos q{,}$ $\rho
_{\sigma }$ and $\tau _{\sigma }$ being the spin-dependent amplitudes for
reflexion and transmission through the impurity region, respectively.
Substituting Eq.~(12) in Eqs.~(9-11) we readily find
\begin{equation}
\tau _{\sigma }=\frac{\left[ 2i(t^{\prime })^{2}/t\right] \sin q}{\left[
2i(t^{\prime })^{2}/t\right] \sin q+\varepsilon \left[ (t^{\prime
}/t)^{2}-1\right] +t_{i}^{2}(\varepsilon _{\sigma }-\varepsilon )^{-1}},
\eqnum{9}  \label{eq4}
\end{equation}
hence the conductance
\begin{eqnarray}
g_{\sigma }=\frac{e^{2}}{h}|\tau _{\sigma }|^{2}=
\frac{e^{2}}{h}\left\{ 1+
\frac{t^{2}}{4(t^{\prime })^{4}\sin q}\left[ \frac{\varepsilon }{t^{2}}%
(t^{\prime \text{ }2}-t^{2})+\frac{t_{i}^{2}}{(\varepsilon _{\sigma
}-\varepsilon )}\right] ^{2}\right\} ^{-1}.  \eqnum{10}  \label{eq9}
\end{eqnarray}

For $t^{\prime }=t$ and $\varepsilon $ equal to the Fermi energy $%
\varepsilon _{F},$ which is the only energy that matters in the linear
response regime, Eq.(14) reduces to the simpler expression\cite{torio2}
\begin{equation}
g_{\sigma }=\frac{e^{2}}{h}\left[ 1+\frac{(\Gamma /2)^{2}}{(\varepsilon
_{\sigma }-\varepsilon _{F})^{2}}\right] ^{-1},  \eqnum{11}
\end{equation}
where $\Gamma =t_{i}^{2}/t=2t_{i}^{2}/\left| v_{F}\right| ,$ with $\left.
v_{F}=d\varepsilon /dq\right| _{q_{F}}$ is the Fermi velocity, and $q_{F}$
the Fermi wave-vector. As a function of $\varepsilon _{\sigma },$ $g_{\sigma
}=0$ if $\varepsilon _{\sigma }=\varepsilon _{F},$ and reaches a value close
to the unitary limit $e^{2}/h$ if $\left| \varepsilon _{\sigma }-\varepsilon
_{F}\right| \gg \Gamma /2.$ The associated valley in the conductance has a
typical width of about $\Gamma .$

On the other hand the mean occupation number on the dot at zero temperature $%
\left\langle n_{\sigma }\right\rangle $ is related to the scattering phase
shift by the Friedel sum rule \cite{hewson,langreth}
\begin{equation}
\left\langle n_{\sigma }\right\rangle =-\frac{1}{\pi }{\rm {Im}\;\ln {(}}%
\varepsilon _{F}-\varepsilon _{\sigma }+i\Gamma /2{),}  \eqnum{12}
\end{equation}
which combined with Eq.(15) yields the relation
\begin{equation}
g_{\sigma }=\frac{e^{2}}{h}\cos ^{2}(\pi \left\langle n_{\sigma
}\right\rangle ).  \eqnum{13}  \label{cos}
\end{equation}
This relation has a geometric origin and actually holds for arbitrary $U$
provided zero temperature limit. This is in contrast to Eqs.\ (13,14,15)
which are applicable only for $U=0$.

We display in the upper panel of Fig. 2 the dot occupancy $\left\langle
n_{\sigma }\right\rangle $ as a function of the resonant level energy $%
\varepsilon _{0}.$ From here on, we set $\varepsilon _{F}=0,$
corresponding to a mean occupancy of one for each of the chain sites.
As
expected, $\left\langle n_{\sigma }\right\rangle $ changes from a value
close to unity for $\varepsilon _{0}\ll 0$, to a value close
to zero for $\varepsilon _{0}\gg 0$, passing through $%
\left\langle n_{\sigma }\right\rangle =1/2$ for $\varepsilon _{0}=0.$ The
lower panel of Fig. 2 corresponds to the conductance $g_{\sigma },$ whose
more noticeable feature is that it is zero for $\varepsilon _{0}=0,$ meaning
that the transmission through the chain is extremely small in this regime.
This is the first example of a Fano resonance discussed in this work; as $%
t_{i}=t=1,$ the width of the resonance $\Gamma /t\simeq 1.$ The symmetric
form of the Fano resonance given by Eq.~(15) is based on the symmetry of the
coupling term in the Hamiltonian (1). Fig. 3 corresponds to the results with
a magnetic field, parametrized by $\Delta /t=1.6.$ The magnetic field breaks
the dot energy degeneracy, $\varepsilon _{\uparrow }\neq \varepsilon
_{\downarrow },$ and the dot occupancy $\left\langle n_{\sigma
}\right\rangle $ depends on spin. The conductance becomes also
spin-dependent, as displayed in the lower panel of Fig. 3. For strong
magnetic fields $\left( \Delta \gg \Gamma \right) ,$ the two Fano resonances
for spin up and spin down electrons are energetically well separated.
Therefore, the current through the channel is completely polarized at $%
\varepsilon _{\uparrow }=\varepsilon _{F}$ and $\varepsilon _{\downarrow
}=\varepsilon _{F}$. For example, for $\varepsilon _{0}=-\Delta /2,$ $%
g_{\uparrow }=0$ while $g_{\downarrow }$ $\simeq e^{2}/h,$ leading to a
perfect spin blocking for electrons with spin $\uparrow .$ The situation
reverses for $\varepsilon _{0}=\Delta /2.$


\begin{figure}[ht]
\includegraphics[width=0.5\textwidth]{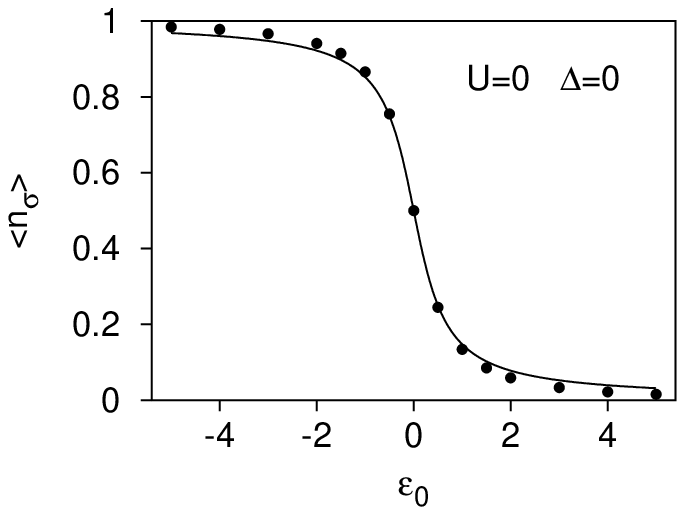}
\includegraphics[width=0.5\textwidth]{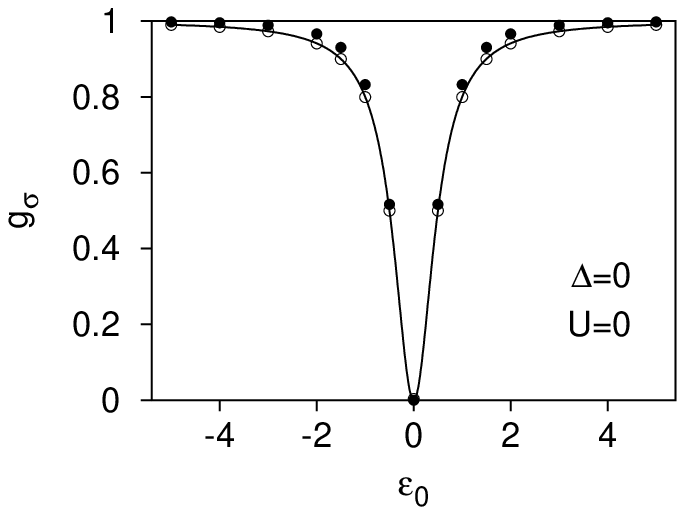}
\caption{Mean dot occupation $\langle n_{\sigma} \rangle$ (top) and
the channel conductance $g_\sigma$ (bottom) in units of $[e^2/h]$ versus the
resonance
energy $\varepsilon_0$ for zero magnetic field and $U=0$. The
results computed numerically by different means are: using
Eq.~(5) (white dots); using $\left\langle n_{\sigma}\right\rangle$ and Eq. (17) (black
dots). Solid line: mean occupation and conductance obtained from the exact
solution by Wiegmann and Tsvelick \cite{WT}. } \label{fig:g}
\end{figure}

\section{Many-body $(U\neq 0)$ Fano resonances}

\subsection{Weak magnetic field $(\Delta \ll k_{B}T_{K})$}

We will first consider the case of zero magnetic field $\Delta =0$ but finite
Coulomb interaction $U$. For zero side-dot/chain coupling $t_{i}=0$, the
single particle excitations on the dot are located at energies $\varepsilon
_{0}$ and $\varepsilon _{0}+U$. When the dot is weakly connected to the
conducting channel (and at very low temperatures, $T<T_{K}$), a resonant
peak in the density of states on the dot arises at the Fermi energy due to
the Kondo effect. By using an approach identical to that of Refs.\ \cite
{glazman,patrick}, we conclude that the channel conductance $g_{\sigma }$ is
controlled entirely by the expectation value of the particle number $\langle
n_{\sigma }\rangle $ as given by Eq.~(17)\cite{torio2}. This reasoning justifies
Eq.~(17)
for any values of $U$ and $\Delta $ \cite{kang-remark}. Note that in the
standard setup of the substitutional configuration the dot's conductance is
proportional to $\sin ^{2}(\pi \langle n_{\sigma }\rangle )$ \cite
{glazman,patrick,langreth} due to the different geometry of the underlying
system.


\begin{figure}[ht]
\includegraphics[width=0.5\textwidth]{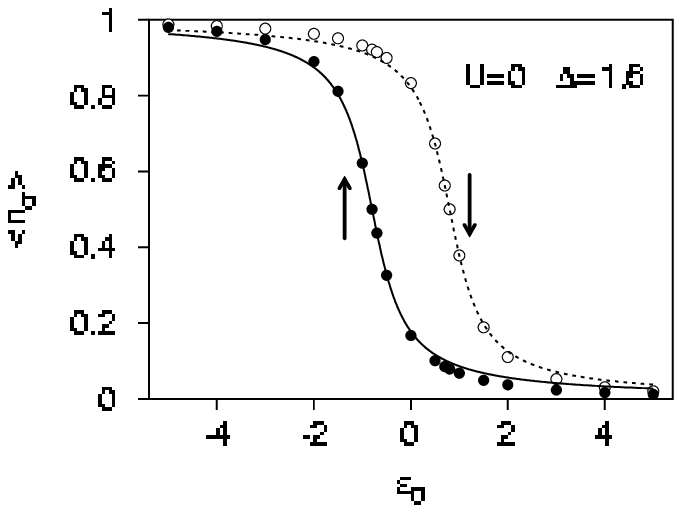}
\includegraphics[width=0.5\textwidth]{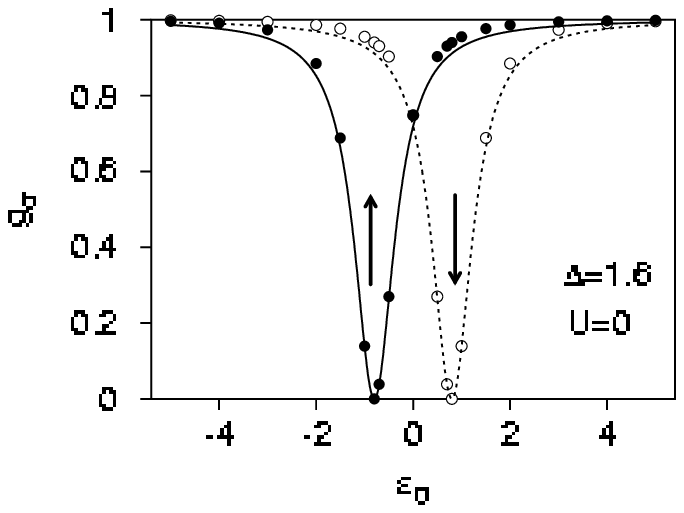}
\caption{Same as Fig. 2 for a finite splitting $\Delta=1.6$ and
$U=0$. The black (white) dots represent the numerical results for
the spin up (spin down) occupation number. The solid and
dotted lines represent the exact results of Eqs.
(16) (upper panel) and (15) (lower panel).}
\end{figure}

In order to compute $\langle n_{\sigma }\rangle $ for finite $U$ we use the
numerical techniques developed in Refs.\ \cite{ferrari,torio1} and
summarized above. We compare our results to those obtained from the exact
solution for $\langle n_{\sigma }\rangle $ by using the Bethe anzats\cite{WT}%
. The latter holds only for the linearized spectrum (which is not a serious
restriction at half-filling) and has a compact form only in the case of zero
magnetic field. We plot in the upper panel of Fig. 4 $\langle n_{\sigma
}\rangle $ versus the resonance level energy $\varepsilon _{0}$ for $U=12$.
In the presence of Coulomb repulsion the mean occupancy of the dot $\langle
n_{\sigma }\rangle $ has a flat region near the value $1/2$ due to many-body
effects. The physical source of the flat region is as follows. For $U\neq 0,$
the condition for near unity dot occupancy per spin (fully occupied dot
regime) is $-(\varepsilon _{0}+U)\gg \Gamma .$ On the other side, the
condition for near zero dot occupancy (empty dot regime) is $\varepsilon
_{0}\gg \Gamma .$ Accordingly, there is a transition regime, characterized
by $0\lesssim \left| \varepsilon _{0}\right| \lesssim U$ where $\langle
n_{\sigma }\rangle $ changes from one to zero. Roughly, this is the size of
the flat region with $\langle n_{\sigma }\rangle \simeq 1/2$ in the upper
panel of Fig. 4. This should be contrasted with the width of this transition
regime in the non-interacting case, which is of the order $\Gamma $ (see
upper panel of Fig. 2).

The flat region around $\langle n_{\sigma }\rangle=1/2 $ for the interacting case
translates inmediatly to a valley of width of order $U$ of small channel
conductance (see lower panel of Fig. 4). Note that similar results for the
limit $U\rightarrow \infty $ have been recently obtained by Bulka and
Stefanski \cite{bulka01}. In their case, however, instead of a valley, the
conductance as a function of $\varepsilon _{0}$ only displays a monotonous
increase from the small conductance regime to the high conductance regime,
as the fully occupied regime is never reached if $U\rightarrow \infty .$ The
conductance found from the numerical evaluation of $\langle n_{\sigma
}\rangle $ for finite $U$ is in better agreement with the exact results \cite
{WT} than the conductance computed from Eq.~(5). Presumably, the discrepancy
originates from finite-size effects (in some cases the Kondo length is larger
than the
system size used in the numerical computation). The numerical calculation of
$\langle n_{\sigma }\rangle $, which involves the energy integration,
appears to be more precise than that of the local density $\rho _{\sigma
}(\varepsilon _{F})$.

\subsection{Strong magnetic field $(\Delta \gg k_{B}T_{K})$}

In the presence of a strong magnetic field such than the Kondo effect is
destroyed $(\Delta \gg k_{B}T_{K}),$ the dot occupation for spin up and spin
down electrons becomes essentially non-overlaping for a large ``window'' of
values of the resonant energies $\varepsilon _{0}$; this is shown in the
upper panel of Fig. 5. The expectation value $\langle n_{\sigma }\rangle $
decreases abruptly from $1$ to $0$ with increasing the resonant energy $%
\varepsilon _{0}$, but it takes the value $1/2$ at different gate voltages
for $\sigma =\uparrow $ and $\sigma =\downarrow $. We will explain this by
using the schematic energy diagram of Fig. 6. Note that due to the dot-chain
hopping $t_{i},$ each dot discrete level should have a width of about $%
\Gamma $. For $\varepsilon _{0}\gg \Gamma ,$ all dot levels are
above the Fermi energy $\varepsilon _{F}(=0),$ the dot is empty, and $%
\langle n_{\uparrow }\rangle \simeq \langle n_{\downarrow }\rangle \simeq 0.$
When decreasing the gate voltage and $\varepsilon _{0}-\Delta /2=\varepsilon
_{F},$ the occupancy of the dot with an electron of spin $\downarrow $ is
allowed, and accordingly $\langle n_{\downarrow }\rangle \neq 0,$ $\langle
n_{\uparrow }\rangle \simeq 0.$ The important point now is that if $U\neq 0,$
the entrance of the second electron with spin $\uparrow $ is blocked until $%
\varepsilon _{0}+U+\Delta /2\simeq \varepsilon _{F},$ that is $\varepsilon
_{0}\simeq -U-\Delta /2.$ The difference in resonant energy between two
events is $U+\Delta $ $(\simeq U).$ On the other side, in the
non-interacting case $U=0,$ and the second electron should only overcome a
much smaller barrier of size $\Delta $ to be in the dot.


\begin{figure}[ht]
\includegraphics[width=0.5\textwidth]{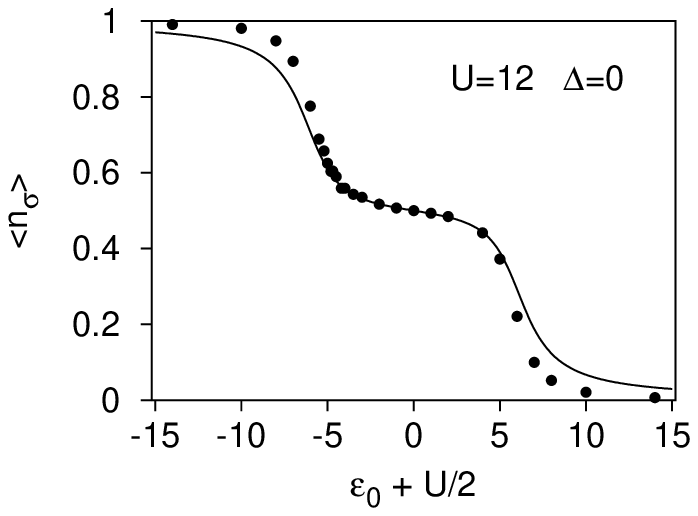}
\includegraphics[width=0.5\textwidth]{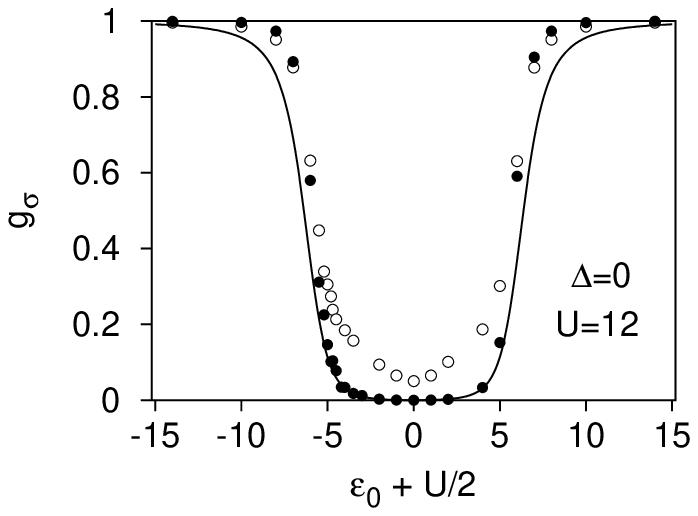}
\caption{Same as Fig. 2 but for $U=12$
}
\end{figure}

Consequently we may obtain a perfect spin filter as in the case of
noninteracting electrons. In Fig.\ 5 (lower panel), we plot the numerical
results for the conductance obtained from Eq.\ (5) for $U/t=12$ with $\Delta
/t=1.6$. Since the dot levels are no longer degenerate the Kondo effect is
totally suppressed even at zero temperature. Indeed no plateaus form near $%
\langle n_{\sigma }\rangle =1/2$ in Fig.\ 5 (upper panel). As a result the
spin-dependent resonances in the conductance have width $\Gamma $. The
effect of large $U$ manifests itself mainly in the mere enhancement of the
level splitting. Using the common terminology of transport through quantum
dots, the results shown in Fig. 5 could be termed as $``$spin-resolved
Coulomb blockade dips ''. At low temperatures $T\ll \Delta $ the mean
occupation number $\langle n_{\sigma }\rangle $ is close to unity while the
occupation number for the opposite spin direction decreases to zero. One can
observe, however, from Fig.\ 5 (lower panel) that there is a mixing of the
levels which gives rise to the non-monotonous behavior of $\langle n_{\sigma
}\rangle $. This results in a small dip of the spin down conductance at the
resonance energy for the spin up electrons and vice versa.

\subsection{Magnetic field quenching of the Kondo-Fano resonance $(\Delta \simeq
k_{B}T_{K})$}

As a reference for our side-dot results for intermediate magnetic fields $%
(\Delta \simeq k_{B}T_{K})$, we start by presenting in Fig. 7 the
conductance results for the geometrical configuration corresponding to Fig.
1a: the {\it substitutional} QD. As discussed above, at zero-temperature and
magnetic field, the Kondo resonance which develops right at the Fermi level
greatly enhances the transmission when the average dot occupancy is close to
1 (Kondo regime). For the symmetric situation $\varepsilon
_{0}=-U/2$, the transmission is perfect and the conductance reaches the
unitary limit $g(0)=2e^{2}/h.$\cite{wiel} This is a highly non-trivial check
from the numerical point of view, and proves that our finite system approach
is capable of sustaining a fully developed Kondo peak with the exact
spectral weight. Within our approach, the Coulomb blockade peaks become
discernible upon magnetic field application. As seen in Fig. 7, with
increasing magnetic field values, spin-fluctuations at the impurity site are
progressively quenched, and the associated enhancement of the conductance
develops towards the usually observed experimentally Kondo valley flanked by
two Coulomb blockade peaks roughly separated by $U$.


The conductance behaviour equivalent to the traditional Kondo effect, corresponding to
the
geometrical arrangement of Fig. 1b, is shown in Fig. 8. In this
configuration, the conductance reaches the unitary limit either when the dot
is fully occupied ($\varepsilon _{0}+U<0$) or empty ($%
\varepsilon _{0}>0$ ). In both cases, the {\it side}
dot weakly perturbs the transmission along the tight-binding chain, as the
possible scattering processes disappear. On the other side, the conductance
becomes progressively blocked as the {\it side} dot enters in the Kondo
regime, reaching the anti-unitary limit $g(0)=0$ exactly at the {\it side}
dot symmetric configuration $\varepsilon _{0}=-U/2.$ In other words, even
though the non-interacting central site provides, in principle,  a channel
for transmission, through its coupling to the {\it side} dot it becomes a
perfectly reflecting barrier.
In Fig. 8b we present the conductance for up spins only (spin down conductance
shows a behaviour symmetric with respect to the Fermi energy). As expected each
channel contributes with $e^2/h$ far from the intermediate valence zone (large
$|\varepsilon_0|$).
Comparing with Fig. 8a, we find that within the low conductance regime, the
spin discriminated conductance reaches zero for one
spin species (up in this case) indicating that for that particular gate voltage the
system transmits only the opposite spin and complete reflects the other. This is
also seen in the local DOS (see Fig. 10b and comment below). By increasing the
magnetic field, and
when the Kondo effect is destroyed completely, the separation between up and down
conductance dips becomes of the order of the dot interaction energy $U$.


\begin{figure}[tbd]
\includegraphics[width=0.5\textwidth]{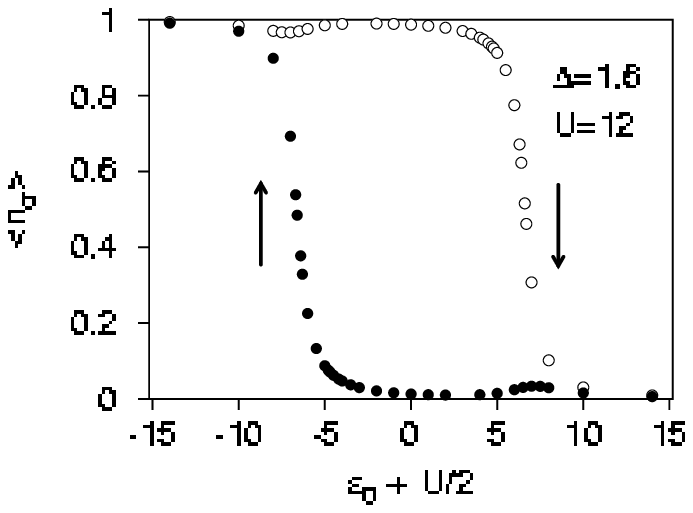}
\includegraphics[width=0.5\textwidth]{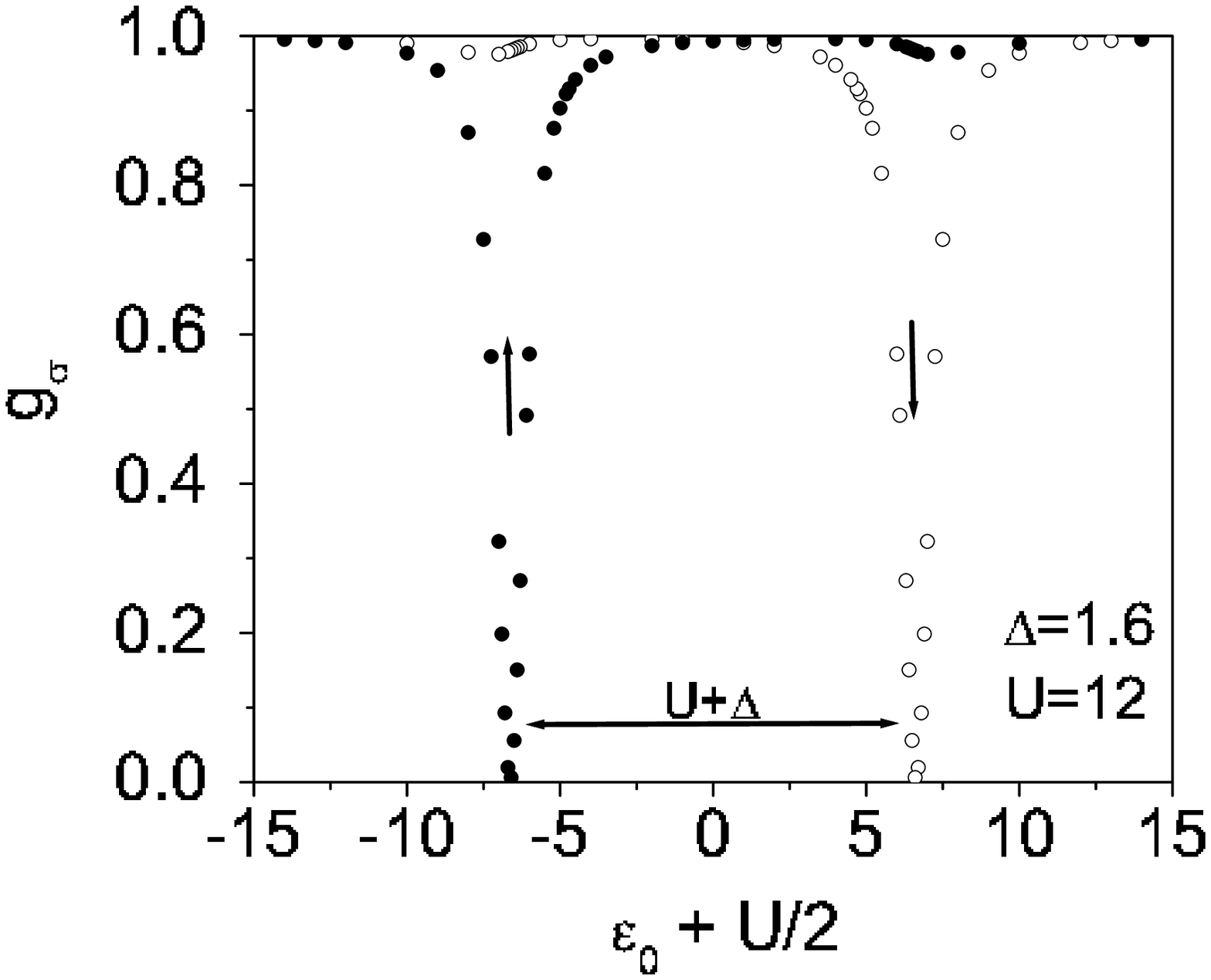}
\caption{Same as Fig. 3 but for $U=12$ }
\end{figure}

The results for the total ($\rho _{\uparrow }+\rho _{\downarrow }$) local
density of states (DOS) shown in Fig. 9 provides a nice qualitative
explanation of the linear conductance results in both geometrical
arrangements. Fig. 9a corresponds to the local DOS at site 0 in the chain,
Fig. 9b to the local DOS at the {\it side} dot. Starting with Fig. 9b, a
well-defined Kondo resonance is discernible around the Fermi level in
absence of magnetic field. The exact (unitary) zero-field result $%
\sum\limits_{\sigma }\rho _{imp,\sigma }(0)=t/(\pi t^{\prime \text{ }%
2})=2/(\pi t)\simeq 0.64$ $($for $t=\sqrt{2}t^\prime $ and $t=1)$ is recovered from our
numerical approach, as discussed above.\cite{hewson}

This local DOS at the impurity site in the {\it side} dot configuration is
equivalent to the local DOS at the impurity in the {\it substitutional} dot
configuration, and gives rise to the unitary limit $g(0)=2e^{2}/h$ discussed
above. From the half-width of the zero-field impurity DOS at half-maximum
(HWHM), we estimate $k_{B}T_{K}/t\simeq 0.15$ for our parameter choice;
note that this estimate qualitatively agrees with the magnetic field values
(Zeeman splittings) for which strong dependences of $g(H)$ with $H$ are
found. In the presence of a magnetic field, the Kondo resonance splits into
two peaks, generating a local minimum between them. As the conductance in
the {\it substitutional} dot configuration is proportional to the dot DOS at
the Fermi level, this explains the abrupt decreasing of $g(H)$ by increasing
$H$ at the middle of the Kondo valley shown in Fig. 7. The most noticeable
feature of Fig. 9a, corresponding to the DOS at site 0 of the chain, is the
profound dip it exhibits around the Fermi level; a pseudo-gap appears at the
symmetric situation $\varepsilon _{0}=-U/2$ just at the Fermi level. The
presence of this pseudo-gap explains the conductance minimum of Fig. 8 at $%
\varepsilon _{0}=-U/2.$ In the presence of a magnetic field, the dip weakens
and accordingly the conductance starts to increase. After a certain
threshold field, the DOS develops a double-well shape around the Fermi
level; the distance between the two well minima is about $2\Delta ,$ which
should be compared with the Zeeman splitting $\Delta .$ If the magnetic
field is big enough $(\Delta /k_{B}T_{K}\gg 1)$, the Kondo effect is
destroyed, the associated low energy coupling between the quantum wire and the side dot
essentially vanishes, and the DOS at site $0$ recovers the semi-elliptical
shape corresponding to the non-interacting chain.

In Fig. 10a we show the local DOS at site $0$ in the chain for spins up only for
different magnetic fields in the symmetric case. It is clearly seen here that for each
field there exists
an energy at which the DOS is zero. For down spins the situation is symmetric with
respect to zero energy. For large fields the Kondo dips disappear and the dips appear
only at the localized energy levels $\varepsilon_{0}-\Delta/2$ and $\varepsilon
_{0}+U+\Delta/2$ (not shown). In Fig. 10b the DOS for up spins at site $0$ for the
asymmetric case and a small magnetic field is shown. We can see here the
evolution of the DOS at the Fermi energy (which defines the behaviour of the
conductance). The up spin DOS minimum evolves towards the Fermi energy as we lower
the gate voltage from the symmetric case until a particular voltage is reached at
which the conductance is zero for this spin species (but finite for the opposite
one).


\begin{figure}[ht]
\includegraphics[width=0.6\textwidth]{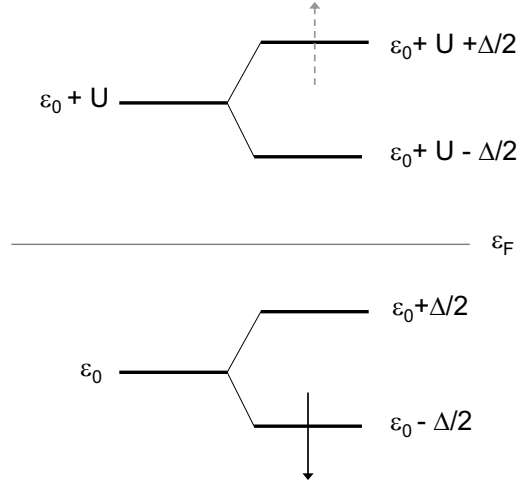}
\caption{Schematic energy diagram of the interacting side-dot. The full down (dashed
up) arrow represents the presence (absence) of an electron with spin down (up).
}
\end{figure}


\begin{figure}[tbd]
\includegraphics[width=0.6\textwidth]{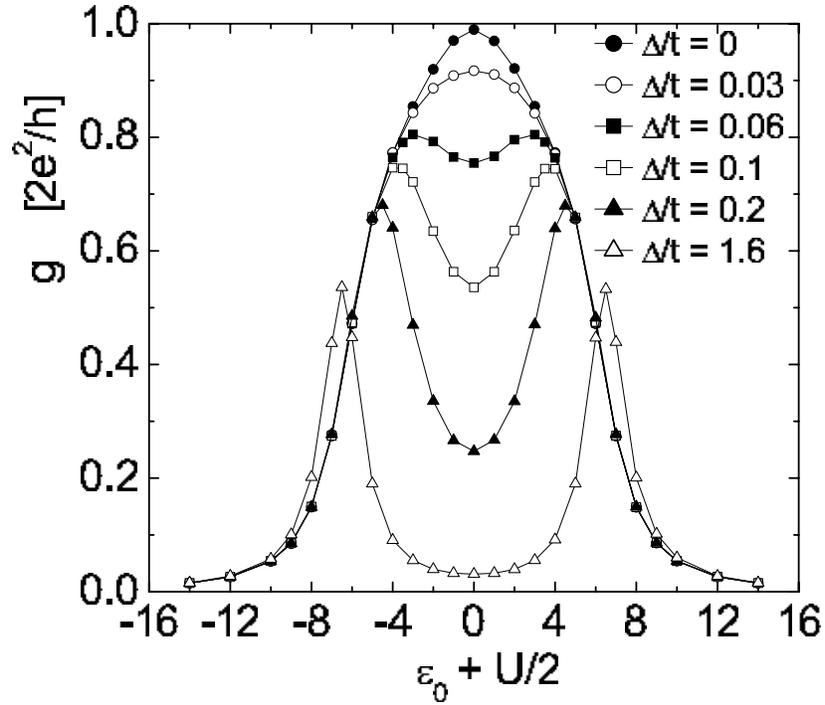}
\caption{Conductance (in units of $[2e^{2}/h]$) for the {\it substitutional} dot
configuration, as a function of impurity level position and for several values
of the magnetic field ($t^{\prime}/t=1/\sqrt{2}$ and $U=12$).
}
\end{figure}

Conceptually, the simplest way to understand these transport features is
using the framework developed by Fano forty years ago.\cite{fano} He
analyzed the properties of a system consisting of a continuous spectrum
degenerate with a discrete level, both non-interacting. Under these
conditions, a dip develops in the density of states of the continuous
spectra, as a result of its interaction with the discrete level. In our case,
the continuous spectra is provided by the thigh-binding chain, while the
role of the discrete level is played by the Kondo peak at the DOS of the
{\it side} dot. As the Kondo effect is destroyed by the magnetic field, the
intrinsic link discussed above between the Kondo peak and its associated
many-body Fano antiresonance implies the weakening of the latter. A most
interesting feature of our calculation is the evolution of $g(H)$ for
increasing $H$: as the Kondo effect is destroyed, the wide minimum develops
a high conductance region around the Fermi level. At high fields, the
conductance shows two dips, separated by $U+\Delta.$ These two dips are
again quite natural in the Fano framework: in this limit the {\it side} dot DOS,
has two single-particle resonances at $\omega=\varepsilon _{0} - \Delta/2$
and $\omega=\varepsilon _{0}+U+\Delta/2$ (not shown in Fig. 9b). They give rise to the
Coulomb blockade peaks in the {\it substitutional} dot configuration: see
Fig. 7 at a high magnetic field situation. However, in the {\it side} dot
configuration, they play the role of two discrete levels which also
interfere destructively with the single-particle chain continuum: the result
is now two Coulomb {\it dips} instead two Coulomb blockade {\it peaks}.
In light of this analysis, it is clear that the wide dip of Fig. 8 results
from the superposition of three quite different dips: a many-body dip around
$\varepsilon _{0}=-U/2$ and two single-particle dips around $\varepsilon
_{0}=0$ and $\varepsilon _{0}+U=0.$ While our calculation is valid for $T=0,$
it seems reasonable to expect similar qualitative behavior by increasing the
temperature: a wide dip (with size $\sim U$) for $T<T_{K},$ two narrow dips
(size $\sim \Gamma$) separated by $U$ for $ T>T_{K}.$


\begin{figure}[tbd]
\includegraphics[ width=0.5\textwidth, height=7.5cm ]{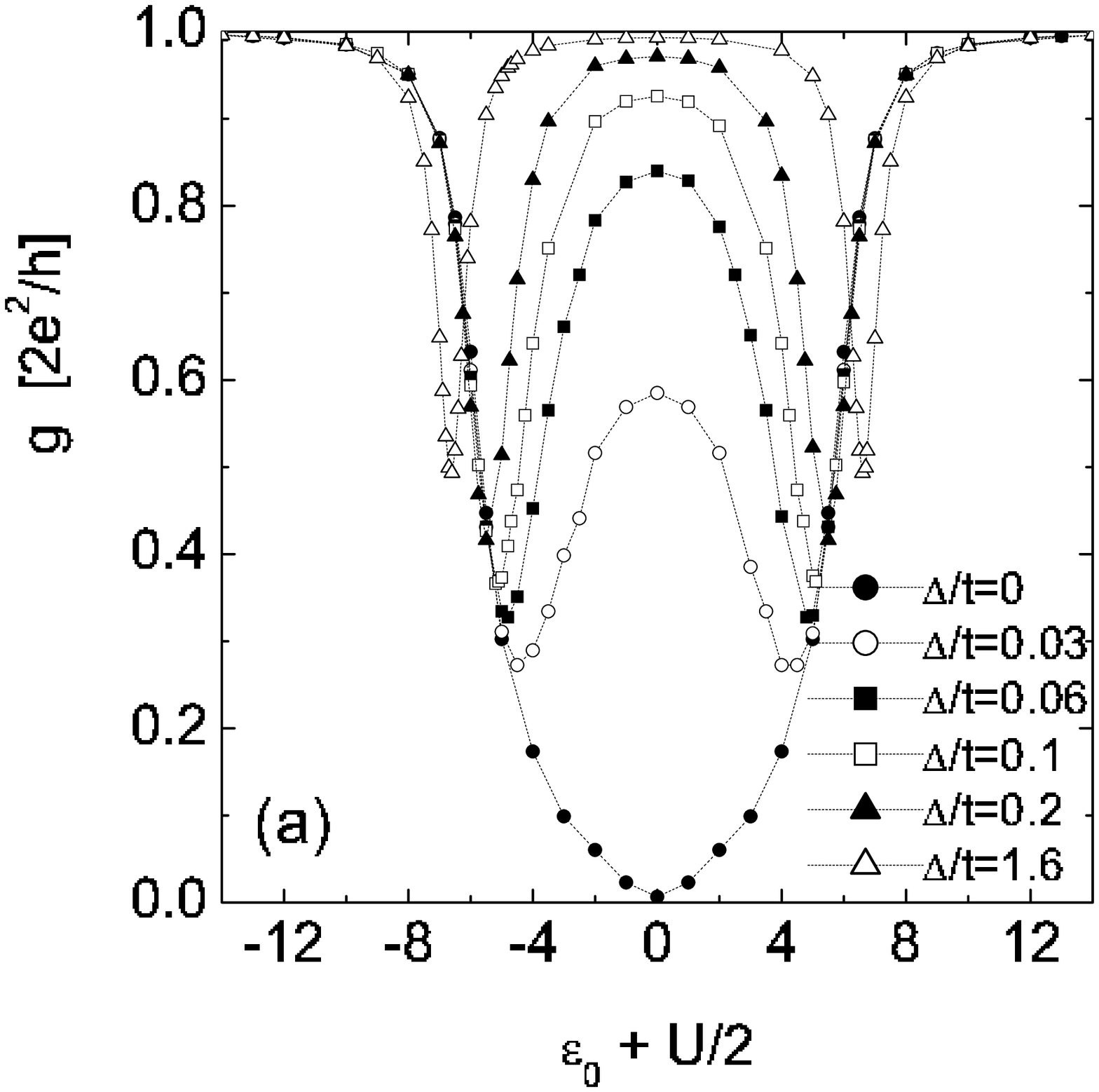}
\includegraphics[ width=0.5\textwidth, height=7.5cm ]{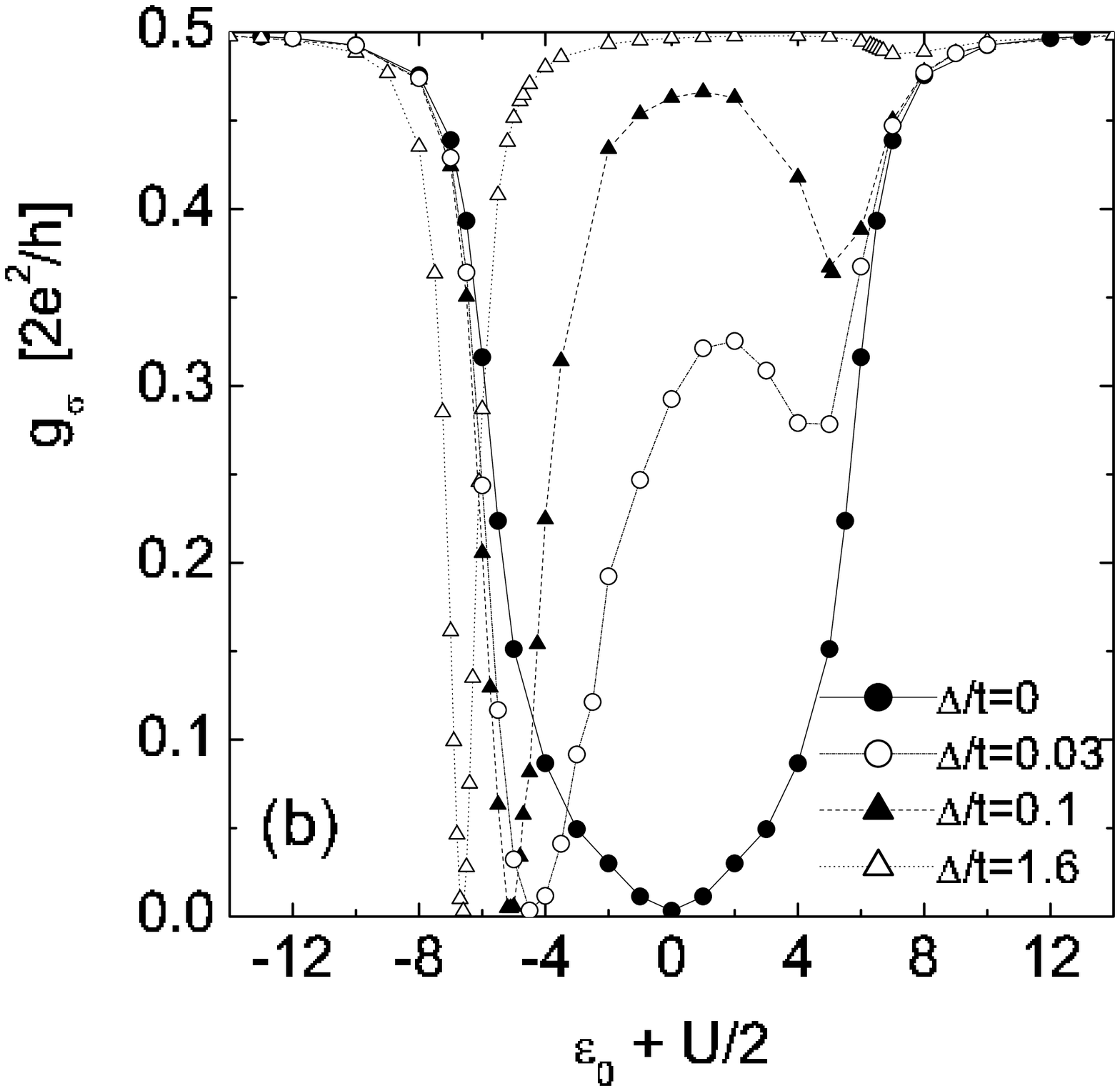}
\caption{Same as Fig. 7 for the {\it side} dot configuration.
a) Total conductance; b) conductance corresponding to the spin up channel (note
the change in units as compared to Fig. 5). Spin
down conductance is symmetric with respect to zero energy. }
\end{figure}


\begin{figure}[ht]
\includegraphics[width=0.5\textwidth]{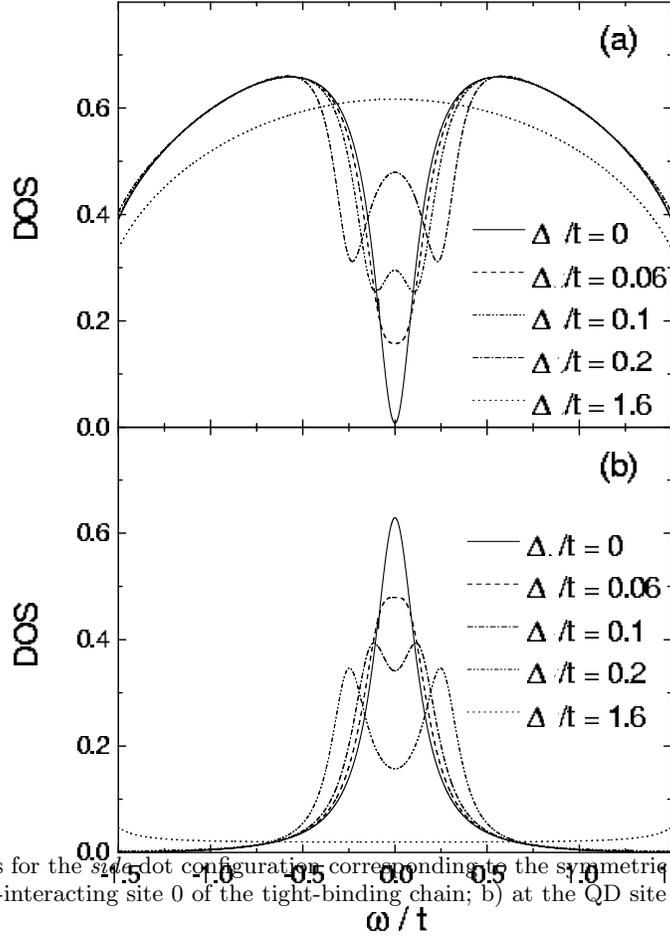}
\vskip -1.5cm \caption{Local density of states for the {\it side}
dot configuration
 corresponding to the symmetric situation $\varepsilon_0 =-U/2$ for different
 magnetic
fields: a) at the
 non-interacting site 0 of the tight-binding chain; b) at the QD site (same
parameters as in Fig. 7).
}
\end{figure}

\section{Conclusions}

We have studied numerically the spin-dependent conductance of electrons
through a single ballistic channel coupled to a {\it side} quantum dot. The
Fano
interference with the dot reverses the dependence of the channel conductance
on the gate voltage compared to that of the {\it substitutional} dot's
conductance. If the spin
degeneracy of the dot levels is lifted, the conductance of the channel
reveals Coulomb blockade dips, or anti-resonances, as a function of the gate
voltage. If the level is spin degenerate the Kondo effect suppresses the
channel conductance in a broad range of the gate voltages.

When the coupling between the channel and the dot is not truly local the
resonances should become highly antisymmetric. In the idealized picture of
the single ballistic channel transport the zero temperature conductance
vanishes at the resonance values of the gate voltage due to the Fano
interference. If there is no spin degeneracy, the resonances have
single-particle origin (for the simple model of Fig. 1b they are separated
roughly by the Coulomb energy $U+\Delta $). At resonance the channel
conductance for a certain spin direction changes abruptly from its ballistic
value $e^{2}/h$ to zero, while the opposite spin conductance remains
approximately at the ballistic value.

This is in contrast with previous calculations that considered a
substitutional QD, where the complementary effect is found, i.e., a perfect
conductance for one spin species. However, in these cases due to the local
correlations and hibridization with the leads, there remains a finite
conductance of the opposite spin species blurring the filtering condition.
In the lateral dot case considered here we find {\it total} suppression of
one spin species, thus leading to a perfect spin filter. Moreover we obtain
that the energetical splitting of the Fano resonances for spin up and spin
down electrons is of the order of the Coulomb interaction $U$ for large $U$
rather than of the order of the Zeeman splitting $\Delta$ as it would be for
noninteracting electrons $U=0$. Thus a strong electron interaction on the
dot (realistic values for $U$ are of the order of 1meV) will allow to apply
very weak magnetic fields (less than 1Tesla) and still observe a significant
splitting of the Fano resonances (provided $k_BT<\Delta$), and thus a high
quality spin filter. This is because not only the spin filter is totally
reflecting electrons with the desired spin $\sigma$, but it will be then
also nearly fully transparent for electrons with opposite spin. Note,
however, that the coupling to the dot should preserve the phase coherence
and there should be only one conducting channel.


\begin{figure}[ht]
\includegraphics[width=0.5\textwidth]{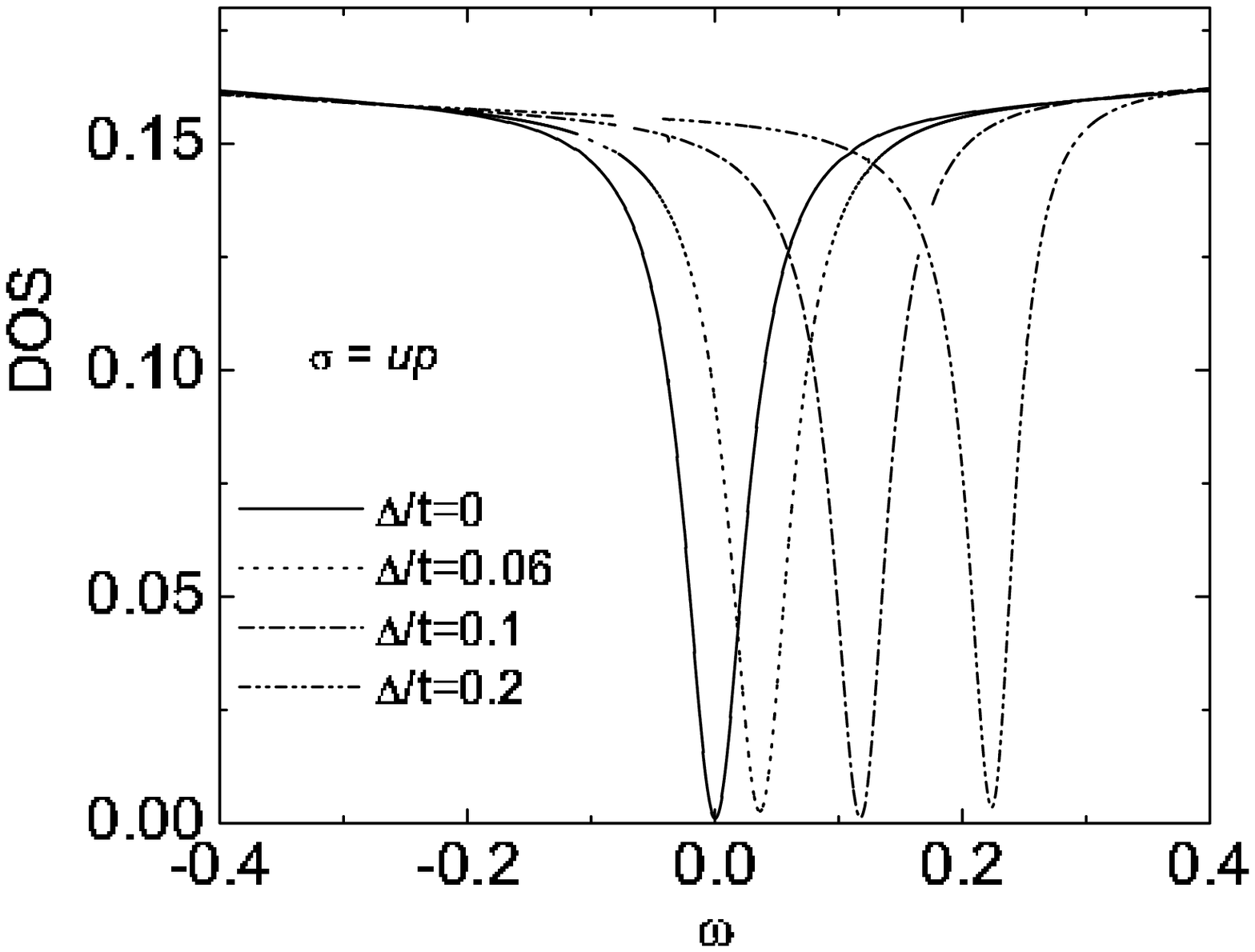}
\includegraphics[width=0.5\textwidth]{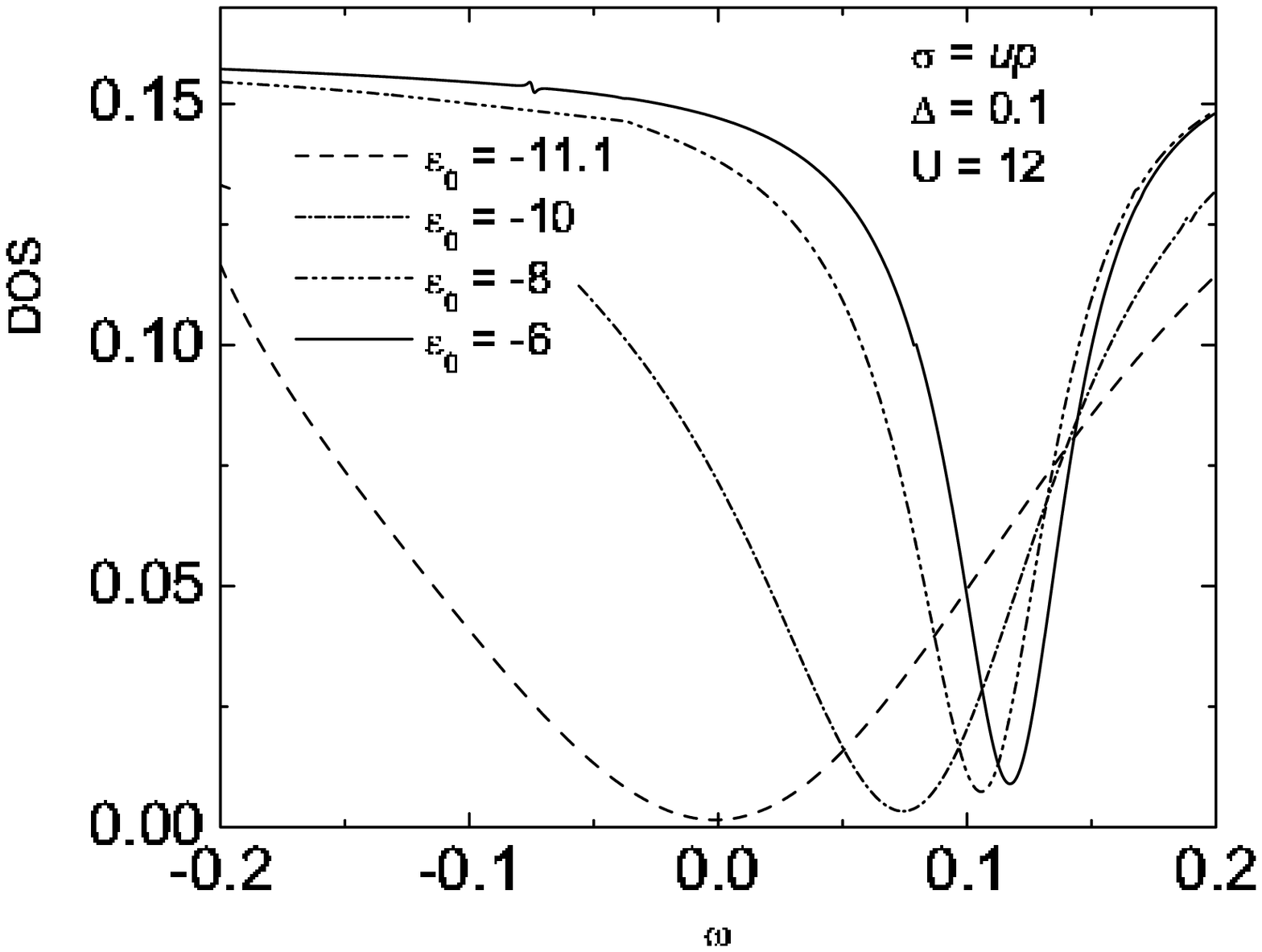}
\caption{Local density of states at site 0 for up spins and the {\it
side} dot configuration
 corresponding to: top: the symmetric situation $\varepsilon_0 =-U/2$ and
different  magnetic fields; bottom: several gate voltage values $\varepsilon_0$
and small magnetic field $\Delta=0.1$.}
\end{figure}

We believe that our calculations also shed light on recent experiments by
G\"{o}res {\it et al.}\cite{gores}{\it \ }Using the same samples of Ref.(%
\cite{david}), and changing the transmission of the left and right tunnel
barriers which connect the dot to the conducting leads, they perform
conductance measurements in the Fano regime (strong coupling leads-dot), the
Kondo regime (intermediate coupling), and the Coulomb blockade regime (weak
coupling leads-dot). The main results concern to the Fano regime, where the
conductance shows asymmetric Fano dips on top of a slowly varying
background. Some features of the dips are reminiscent of the Kondo effect,
as the temperature dependence of the dips amplitudes (Fig. 4 in Ref. \cite
{gores}), and others of the Coulomb blockade effect, as the typical
diamond-shaped structure in differential conductance measurements (Fig. 5 in
Ref. \cite{gores}). Both features are easily explained by our results. For
example, from Fig. 8a, and considering that finite temperature weakens the
Kondo
resonance at the Fermi energy  in a similar way as the magnetic field, one can
expect a strong dependence of dip
amplitude with temperature, with the dip amplitude decreasing for increasing
temperature. Besides, the behavior of dips in differential conductance
measurements should be completely analogous to the related Coulomb blockade
peaks, leading to diamond-shaped structures for the dips. Our results also
shed light on the somehow related problem of the persistent current in a
mesoscopic ring with a {\it side} dot.\cite{affleck,eckle} The results
remain controversial on this issue, as Ref. \cite{affleck} found a detrimental
effect of the {\it side} dot on the persistent current when the Kondo effect
is operative. An opposite result was found in Ref. \cite{eckle}, with the
ring exhibiting a perfect (unitary) persistent current in the Kondo regime.
Our results for the open configuration of the present work provide naturally
strong support to the enhancement effect found by Affleck and Simon.\cite
{affleck}

The authors acknowledge CONICET for support. We would like to thank J. Simonin, M.
Titov, S. Flach and A. Miroshnichenko for useful comments and discussions.
Part of this work was done under projects Fundaci\'{o}n Antorchas 14116-168 and
14116-212.


\begin{references}
\bibitem{kang-remark}  For a detailed derivation in the absence of magnetic
fields we refer to Ref.~\cite{kang}.
\end{references}
\end{document}